\documentclass[prd,superscriptaddress,amsfonts,amssymb,amsmath,showpacs,twocolumn,floatfix,nofootinbib]{revtex4-2}
\usepackage{bm}
\usepackage{amsfonts}
\usepackage{latexsym}
\usepackage[latin1]{inputenc}
\usepackage{graphicx}
\usepackage{amsmath}
\usepackage{palatino}
\usepackage{mathpazo}
\usepackage{textcomp}
\usepackage{float}
\usepackage{booktabs}
\usepackage{dcolumn}
\usepackage{ragged2e}
\usepackage{hyperref}
\hypersetup{colorlinks,citecolor=green}
\hypersetup{colorlinks=true,linkcolor=blue,filecolor=blue,    urlcolor=magenta}
\usepackage{amsmath}
\usepackage{xcolor}
\usepackage[caption=false]{subfig}
\usepackage{commath}

\allowdisplaybreaks[1]

\begin{document}

\color{black}       

\title{Properties of interacting quark star in light of Rastall gravity}

\author{Ayan Banerjee} \email[]{ayanbanerjeemath@gmail.com}
\affiliation{Astrophysics Research Centre, School of Mathematics, Statistics and Computer Science, University of KwaZulu--Natal, Private Bag X54001, Durban 4000, South Africa}

\author{Anirudh Pradhan}
\email{pradhan.anirudh@gmail.com}
\affiliation{Centre for Cosmology, Astrophysics and Space Science, GLA University, Mathura-281 406, Uttar Pradesh, India}

\author{\.{I}zzet Sakall{\i}}\thanks{\textcolor{blue}{Corresponding author}}
\email{izzet.sakalli@emu.edu.tr}
\affiliation {Physics Department, Eastern Mediterranean University, Famagusta 99628, North Cyprus via Mersin 10, Turkey.}

\author{Archana Dixit}
\email{archana.ibs.maths@gmail.com}
\affiliation {Department of Mathematics,
Gurugram University, Gurugram-122003, Haryana, India.}


\date{\today}

\begin{abstract}
This study explores the properties of quark stars (QS) formulated with an interacting quark matter equation of state (EoS) within the framework of Rastall gravity, a modified theory of gravity. We derive the mass-radius relationships and calculate the maximum gravitational masses and their corresponding radii, comparing these results under both Rastall gravity and general relativity. Our analysis incorporates recent observational data, including the GW190425 event, to constrain the model parameters ($\bar{\lambda}, \eta, B_{\rm eff}$). We also assess the stability of these quark stars by evaluating their static stability, adiabatic index, and sound velocity profiles, \textcolor{black}{thus confirming their viability within the modified gravitational framework}.

\end{abstract}

\maketitle

\section{Introduction}
\textcolor{black}{Recent advancements in astrophysical observations have greatly
enhanced our understanding of compact stellar objects, especially QSs and neutron stars (NSs) \cite{isZhang:2024xod,isSu:2024znh,isKolomeitsev:2024gek,isPretel:2024lae,isSen:2024reu,isFarrell:2024bka,isTANGPHATI2024}.} These observations are crucial, as they provide stringent constraints on the EoS of dense matter under extreme conditions that cannot be replicated in terrestrial laboratories \cite{isBaym:2017whm,isLOFT:2011pkp}. In this paper, we focus on the theoretical modeling and implications of the QSs formed by \textcolor{black}{interacting quark matter} (IQM) within the framework of Rastall gravity -- a modified theory of gravity \cite{isDarabi:2017coc,isOliveira:2015lka,isHeydarzade:2016zof,isFabris:2012hw,isOvgun:2019jdo}. This theory connects the energy-momentum tensor directly to the spacetime curvature, in contrast to  \textcolor{black}{general relativity (GR)}, which does not permit a non-zero divergence of the tensor \cite{isDeMoraes:2019mef}. The exploration of QSs, which are hypothesized to be composed of deconfined quark matter, hinges critically on the EoS used to describe them \cite{isAnnala:2017tqz,isKurkela:2014vha,isMariani:2019vve}. Traditionally, quark matter was modeled using simplistic bag models which encapsulate the effects of a vacuum energy density (the so-called bag constant) but ignore the interactions among quarks 
\cite{isFreedman:1977gz,isHasenfratz:1977dt,isBaym:1976yu}
. However, recent models, informed by perturbative Quantum Chromodynamics (pQCD) \cite{isReya:1979zk,isGross:2022hyw}, suggest significant modifications to the EoS when interactions and color superconductivity are taken into account \cite{isBuballa:2003qv}. These advanced models predict that quark matter could be absolutely stable under certain conditions, and thus, compact stars made entirely of quark matter (or containing significant quark cores) might exist in the universe \cite{isCollins:1974ky,isAnnala:2019puf,isWeissenborn:2011qu}. 

The presence of massive pulsars such as PSR J0348+0432, PSR J0740+6620, and the recent GW190814 event, which involved a compact object of about 2.6 solar masses, potentially a massive neutron star or a light black hole, suggests that some NSs could be massive QSs \cite{isMiller:2021qha}. These observations challenge traditional NS models based on nucleonic matter and support scenarios where significant fractions of the star's core might consist of deconfined quarks \cite{isMcLerran:2007qj,isGhaemmaghami:2023sho,isMishra:2024jmc}. Given this backdrop, theories of modified gravity, in particular the Rastall gravity \cite{isShankaranarayanan:2022wbx} (and references therein), offer intriguing alternatives for studying the structure of compact stars. Rastall's theory extends Einstein's GR by modifying the conservation law of the energy-momentum tensor \cite{isYuan:2016pkz}. This modification suggests a dynamic coupling between matter and spacetime geometry, which is not predetermined but varies with the curvature of spacetime. This feature makes it a suitable candidate for investigating the properties of dense matter under extreme gravitational fields \cite{isAntoniadis:2013pzd,isCardoso:2019rvt,isBaiotti:2016qnr,issakalli:2022xrb}.

{\color{black}

In recent years, hybrid star models, which combine hadronic and quark matter, have emerged as viable candidates for explaining massive NSs, especially when considering phase transitions under extreme conditions \cite{PhysRevD109063008}. These models, along with hyperon-inclusive EoSs, address challenges like the "hyperon puzzle," which arises from the softening effect of hyperons that can reduce the maximum mass of NSs \cite{EurPhysJA5620}. Studies indicate that adjusting parameters such as vector interactions or incorporating strong magnetic fields can produce stiffer EoSs, allowing NSs to reach observed masses above $2 M_\odot$ \cite{PhysRevC105015807,PhysRevC107045804}. Our current study on QSs under Rastall gravity focuses solely on IQM, yet future work could extend to hybrid or hyperonized matter models. This expansive methodology would facilitate a more thorough comparison of EoSs, assessing the reliability of Rastall gravity in explaining compact stars with various makeups.
}

In this study, we consider the IQM in the context of compact stars and analyze them within the Rastall gravity framework. We shall discuss how Rastall gravity modifications affect the mass-radius relationship of the QSs and their stability. To this end, we first investigate the properties of QSs, modeled using an IQM EoS within the Rastall gravity. Then, we derive the mass-radius relationships, calculating the maximum gravitational masses and their corresponding radii, and comparing these findings under both Rastall gravity and GR. By incorporating recent observational data, including the GW190425 event, we constrain model parameters ($\bar{\lambda}, \eta, B_{\rm eff}$) and evaluate the stability of QSs \cite{isTangphati:2024war} through analyses of static stability, adiabatic index, and sound velocity profiles, confirming their viability in the modified gravitational context. \textcolor{black}{It is also worth noting that the parameter $\eta$ in Rastall gravity, which measures the deviation from GR, has been explored across a wide range in various astrophysical studies \cite{isOliveira:2015lka,isHeydarzade:2016zof,isMaulana2019,isTANGPHATI2024}. Values such as $\eta=0.83$ have been analyzed to assess how modified gravity theories like Rastall's might influence the properties of compact objects or cosmic structure without conflicting with existing observational data. However, tighter constraints on $\eta$ may emerge with advancements in gravitational wave observations and precision astrophysical measurements.}

The organization of this paper is as follows. In Section \ref{sec2}, we introduce the EoS for strange quark matter (SQM), emphasizing the effects of interactions and color superconductivity derived from the pQCD. Section \ref{sec3} details the theoretical framework of Rastall gravity, presenting the modified Tolman-Oppenheimer-Volkoff (TOV) equations adapted for this theory. We apply these equations to study the structural properties of QSs. Numerical results and discussions on the mass-radius relationship and the impact of various parameters like the Rastall coupling constant $\eta$, the interaction parameter $\bar{\lambda}$, and the effective bag constant $B_{\rm eff}$ are provided in Section \ref{sec4}. Section \ref{sec5} investigates the stability of these QS configurations through several physical criteria, including static stability, adiabatic index, and sound velocity profiles. We conclude in Section \ref{sec6} with a summary of our findings and their implications for the physics of compact stars and future observational tests. 

\section{E\lowercase{o}S of SQM} \label{sec2}

As mentioned in the introduction, the authors in \cite{isZhang:2020jmb} have studied the effects of QCD interactions by considering interquark effects derived from pQCD corrections and the phenomenon of color superconductivity. Utilizing this newly proposed IQM EoS, we appraise the structural properties
of this star within the framework of Rastall gravity theory.

Following \cite{isZhang:2020jmb,isZhang:2021fla}, we have determined that the density $(\rho)$ for IQM is related to the isotropic pressure $(P_r)$ through the following relationship:
{\fontsize{9.7}{12}
\begin{align}
p=\frac{1}{3}(\rho-4B_{\rm eff})+ \frac{4\lambda^2}{9\pi^2}\left(-1+{\rm sgn}(\lambda)\sqrt{1+3\pi^2 \frac{(\rho-B_{\rm eff})}{\lambda^2}}\right),
\label{eos_tot}
\end{align}}
where the effective bag constant $B_{\rm eff}$ accounts for the nonperturbative contribution from the QCD vacuum.
\begin{eqnarray}
\lambda=\frac{\xi_{2a} \Delta^2-\xi_{2b} m_s^2}{\sqrt{\xi_4 a_4}},
\label{lam}
\end{eqnarray}
where $m_s$ denotes the strange quark mass, and $\Delta$ is the gap parameter. The coefficient $a_4$ parameterizes the QCD corrections from one-gluon exchange for gluon interaction at $O(\alpha_s^2)$, and varies from small values up to $a_4=1$. The sign of $\lambda$, represented by ${\rm sgn}(\lambda)$, is positive as long as $\Delta^2/m_s^2 > \xi_{2b}/\xi_{2a}$. The constant coefficients in $\lambda$ are
\begin{align} \label{izzetxi}
(\xi_4,\xi_{2a}, \xi_{2b}) = \left\{ \begin{array} {ll}
\bigg(\big( \left(\frac{1}{3}\right)^{\frac{4}{3}}+ \left(\frac{2}{3}\right)^{\frac{4}{3}}\big)^{-3},1,0\bigg), & \textrm{2SC phase,}\\
(3,1,3/4), & \textrm{2SC+s phase,}\\
(3,3,3/4),&   \textrm{CFL phase,}
\end{array}
\right.
\end{align}
and they characterize the possible phases of color superconductivity. As noted in Ref. \cite{isZhang:2020jmb}, we now introduce the dimensionless rescaling:
\begin{align}
\bar{\rho}=\frac{\rho}{4B_{\rm eff}}, \,\, \bar{p}=\frac{p}{4B_{\rm eff}},  \,\,
\label{rescaling_prho}
\end{align}
and 
\begin{align}
 \bar{\lambda}=\frac{\lambda^2}{4B_{\rm eff}}= \frac{(\xi_{2a} \Delta^2-\xi_{2b} m_s^2)^2}{4B_{\rm eff}\xi_4 a_4}.
 \label{rescaling_lam}
\end{align}
By applying the rescalings (\ref{rescaling_prho}) and (\ref{rescaling_lam}), we derive the dimensionless form of Eq.~(\ref{eos_tot}), which is expressed as follows:
\begin{align}
\bar{p}=\frac{1}{3}(\bar{\rho}-1)+ \frac{4}{9\pi^2}\bar{\lambda} \left(-1+{\rm sgn}(\lambda)\sqrt{1+\frac{3\pi^2}{\bar{\lambda}} {(\bar{\rho}-\frac{1}{4})}}\right).
\label{eos_p}
\end{align}
Interestingly, Eq. (\ref{eos_p}) simplifies to $\bar{p}=\frac{1}{3}(\bar{\rho}-1)$ when $\bar{\lambda} \to 0$, resembling the rescaled conventional noninteracting quark matter EoS. It is noted from Eq. (\ref{eos_p}) that extremely large positive values of $\bar{\lambda}$ yield
\begin{align}
\bar{p}\vert_{\bar{\lambda}\to \infty}=\bar{\rho}-\frac{1}{2}. 
\label{eos_infty1}
\end{align}
or, equivalently, $p={\rho}-2B_{\rm eff}$ after scaling back using Eq. (\ref{rescaling_prho}). However, for negative values of $\lambda$, taking the EoS to the $\bar{\lambda} \to \infty$ limit, Eq. (\ref{eos_p}) does not yield a finite form. As expected, the positive increasing values of $\bar{\lambda}$ correlate with the maximum mass of QS, as detailed in Refs. \cite{isZhang:2020jmb,isZhang:2021fla}.

It is worth noting that the coefficients $\xi$ seen in Eq. \ref{izzetxi} correspond to different phases of quark matter: the two-flavor superconducting (2SC) phase, the 2SC+s phase (which includes strange quarks), and the color-flavor-locked (CFL) phase, where all three quark flavors pair symmetrically. In our study, we focus on the CFL phase, as it is expected to occur at the highest densities relevant to the core of QSs. The CFL phase provides a stiffer EoS, allowing for higher maximum masses, which is essential for comparing our models with observed massive NSs and the compact object from GW190814. This choice ensures that we explore the most extreme cases of QSs under Rastall gravity. Moreover, we use specific values for the gap parameter $\Delta$ and the strange quark mass $m_s$. For the CFL phase, we adopt $\Delta \approx 100$ MeV, a typical value used in studies of color superconductivity, which provides a sufficiently stiff EoS for modeling massive QSs. The strange quark mass is set to $m_s \approx 150$ MeV, consistent with the effective mass of the strange quark in high-density quark matter. These values ensure that our results align with established research in quark matter EoS modeling.

\section{Rastall Gravity Theory and Modified TOV Equations}\label{sec3}

As discussed in the introduction, Rastall gravity was proposed as a simple modification of GR \cite{isRastall:1972swe,isRastall:1976uh}. According to Rastall's proposal, the covariant divergence of the energy-momentum tensor is non-vanishing and proportional to the curvature scalar, i.e., $T^{\mu}_{\nu;\mu} \propto R_{;\nu}$. This assumption leads to a violation of the usual conservation laws. Consequently, we derive the Rastall field equation as \cite{isRastall:1972swe,isRastall:1976uh} (for more information, see \cite{isOliveira:2015lka}) 
\begin{equation}
    G_{\mu \nu} - \gamma g_{\mu\nu} R = 8 \pi G T_{\mu\nu}, 
\label{Rastall_field}
\end{equation}
where $\gamma \equiv \left( \eta - 1 \right)/2$ and $\eta$ is a free parameter known as Rastall's coupling constant, which measures the deviation from GR. Notably, when $\eta=1$, the framework of GR is recovered. The conservation law of the energy-momentum tensor, adjusted to align with Rastall's theory, is given by:
\begin{eqnarray}
    \nabla^{\mu} T_{\mu \nu} = \frac{1}{2} \left( \frac{\eta - 1}{2\eta - 1} \right) \nabla_{\nu} T \neq 0. 
    \label{non_conserved}
\end{eqnarray}
This expression diverges at $\eta = 1/2$, a value we avoid in our analysis. Rastall's theory can thus be seen as a nonminimal coupling between matter and geometry. For simplicity, we rewrite Eq. (\ref{Rastall_field}) as:
\begin{eqnarray}
    G_{\mu \nu} = 8 \pi \tau_{\mu \nu}, \label{ref9}
\end{eqnarray}
where $\tau_{\mu \nu}$ represents the effective energy-momentum tensor, defined as:
\begin{eqnarray}
    \tau_{\mu \nu} = T_{\mu \nu} - \frac{1}{2} \left( \frac{\eta - 1}{2\eta - 1} \right) g_{\mu \nu} T.
    \label{T_eff}
\end{eqnarray}

To analyze the QSs within Rastall gravity, we consider a static and spherically symmetric spacetime, simplifying our calculations. The metric ansatz is given by:
\begin{eqnarray}
	ds^2 = -e^{\nu(r)} dt^2 + e^{\lambda(r)} dr^2 + r^2 \left( d\theta^2 + \sin^2 \theta d\phi^2 \right).
	\label{line_element}
\end{eqnarray}
Here, $\nu(r)$ and $\lambda(r)$ are arbitrary functions of the radial coordinate. We assume an isotropic perfect fluid for the energy-momentum tensor:
\begin{eqnarray}
T_{\mu \nu} = (\rho+p)u_\mu u_\nu + p g_{\mu \nu}, 
\label{eq3}
\end{eqnarray}
where $u^\mu$ is the four-velocity of the fluid, with $\rho(r)$ and $p(r)$ representing the density and pressure, respectively. The effective energy-momentum tensor, $\tau_{\mu \nu}$, combines the usual stress-energy tensor profile:
\begin{eqnarray}
    \rho_{\rm{eff}} &=& \frac{1}{2} \left( \frac{3\eta - 1}{2\eta - 1} \right) \rho + \frac{3}{2} \left( \frac{\eta - 1}{2\eta - 1} \right) p, 
    \label{rhotilde} \\
    p_{\rm{eff}} &=& \frac{1}{2} \left( \frac{\eta - 1}{2\eta - 1} \right) \rho + \frac{1}{2} \left( \frac{\eta + 1}{2\eta - 1} \right) p. 
    \label{ptilde} 
\end{eqnarray}

Implementing Eqs. (\ref{ref9}) and (\ref{line_element}), we derive the Rastall version of the modified TOV equations as follows (using natural units where $G = c = 1$):
\begin{eqnarray}
    \frac{d p_{\rm{eff}}}{dr} &=& - \frac{\left( \rho_{\rm{eff}} + p_{\rm{eff}} \right) \left( m_{\rm{eff}} + 4 \pi r^3 p_{\rm{eff}} \right)}{r^2 \left( 1 - \frac{2 m_{\rm{eff}}}{r} \right)}, 
    \label{dpdr} \\
    \frac{d m_{\rm{eff}}}{dr} &=& 4 \pi r^2 \rho_{\rm{eff}},
    \label{dMdr}
\end{eqnarray}
where $m_{\rm{eff}}$ signifies the effective mass. The metric potentials are related as described in \cite{Velten:2016bdk,Maulana:2019}:
\begin{eqnarray}
    \frac{d \nu}{dr} &=& - \frac{2}{\rho_{\rm{eff}} + p_{\rm{eff}}} \frac{d p_{\rm{eff}}}{dr}, \\
    e^{-\lambda} &\equiv& 1 - \frac{2 m_{\rm{eff}}}{r}.
\end{eqnarray}

In \cite{Velten:2016bdk}, the mass function is defined as:
\begin{eqnarray}
    m &\equiv& \int 4\pi r^2 \rho(r) dr.
\end{eqnarray}

To establish a relationship between the effective mass $m_{\text{eff}}$ and the mass $m$, we start by noting that $m(r)$ is defined as the integral of the standard energy density $\rho$, while $m_{\text{eff}}(r)$ is defined in terms of the effective energy density $\rho_{\text{eff}}$. Similar to the relation between $(p, \rho)$ and $(p_{\text{eff}}, \rho_{\text{eff}})$, we have:

\begin{equation} m(r) = \int_0^r 4\pi r'^2 \rho(r') , dr', \end{equation}

\begin{equation} m_{\text{eff}}(r) = \int_0^r 4\pi r'^2 \rho_{\text{eff}}(r') , dr'. \end{equation}

Using the relation between $\rho_{\text{eff}}$ and $\rho$, the effective mass can be expressed as:

\begin{eqnarray} 
m_{\text{eff}}(r)&= \frac{1}{2} \left( \frac{3\eta - 1}{2\eta - 1} \right) m(r) +\frac{3}{2} \left( \frac{\eta - 1}{2\eta - 1} \right) \\ \notag
& \times \int_0^r 4\pi r'^2 p(r') , dr'. 
\end{eqnarray} 

This equation relates $m_{\text{eff}}$ to $m$ and includes a contribution from the pressure term modified by the parameter $\eta$, similar to the relationships for $(p, \rho)$ and $(p_{\text{eff}}, \rho_{\text{eff}})$.

For a complete description of star solutions, the modified TOV equations (\ref{dpdr}) and (\ref{dMdr}) must be supplemented by an EoS for the fluid. In this context, we numerically solve the differential equations using the EoS given in Eq. (\ref{eos_p}) and compute the mass-radius relation for various model parameters. Proper boundary conditions are essential for numerical solutions:
\begin{align}
   \rho(0) = \rho_c,    ~~~~    m(0) = 0,
   \label{BC_Inner}
\end{align}
ensuring regularity at the stellar interior with $\rho_c$ representing the central energy density. To ensure consistency in our analysis, the initial condition for the effective mass $m_{\text{eff}}$ at the center of the star is given by $m_{\text{eff}}(0) = 0$. This reflects the expectation that the enclosed mass at the center is zero, aligning with the differential equation for $m_{\text{eff}}$ presented in Eq. \eqref{dMdr}. While we specify $\rho(0) = \rho_c$ as the initial condition for the central density, this is used to determine the initial effective pressure $p_{\text{eff}}(0)$ via the EoS and the relation between $p_{\text{eff}}$ and $\rho$. The effective pressure is given by Eq. \eqref{ptilde}, which is derived from the EoS. Namely, the correct boundary conditions are applied for solving the differential equations for $p_{\text{eff}}$ and $m_{\text{eff}}$. This adjustment ensures that the boundary conditions and the governing equations are consistent throughout the analysis.

We start integration from the center with $\rho_c$ and continue towards the surface where the pressure vanishes, i.e., $p(r_{\rm s}) = 0$. Here, $r_{\rm s}$ denotes the star's radius. We also assume the Schwarzschild solution as the exterior solution for the star:
\begin{eqnarray}
e^{-\lambda} = 1 - \frac{2M}{r_s}, \label{e11}
\end{eqnarray}
with $m(r_{\rm s}) \equiv M$ being the total mass of the star. In this exterior solution, $\lambda$ is not a constant, but a function of $r$ that decreases as $r$ increases from the star's surface outward. The form of $e^{-\lambda}$ corresponds to the standard Schwarzschild metric, where the curvature falls off with $1/r$, reflecting the mass $M$ of the star. This ensures that the spacetime outside the star is asymptotically flat, consistent with the vacuum solution.

\begin{figure}[h]
    \centering
    \includegraphics[width=8cm, height=6.3cm]{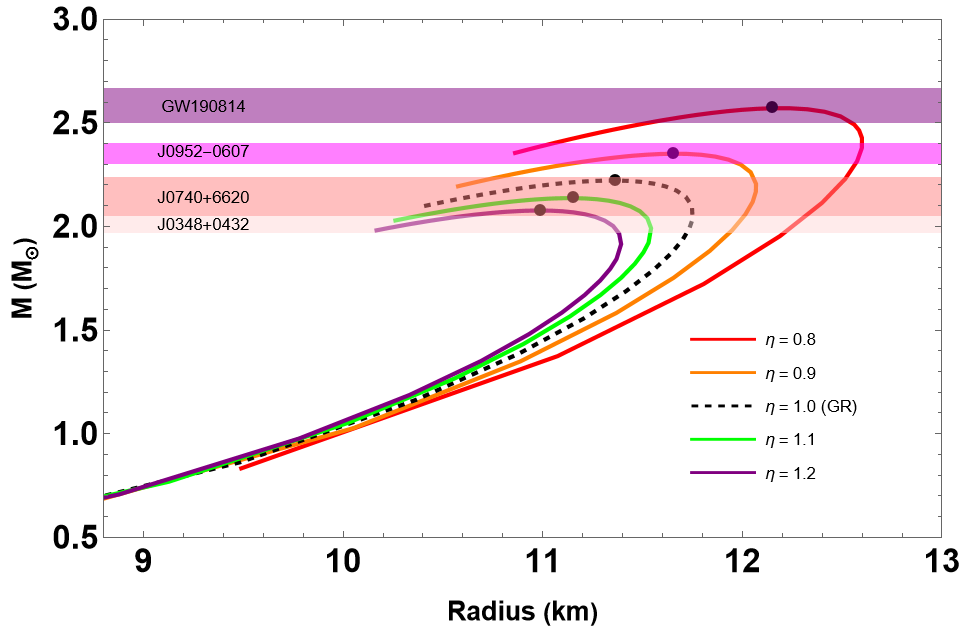}
    \includegraphics[width=8cm, height=6.3cm]{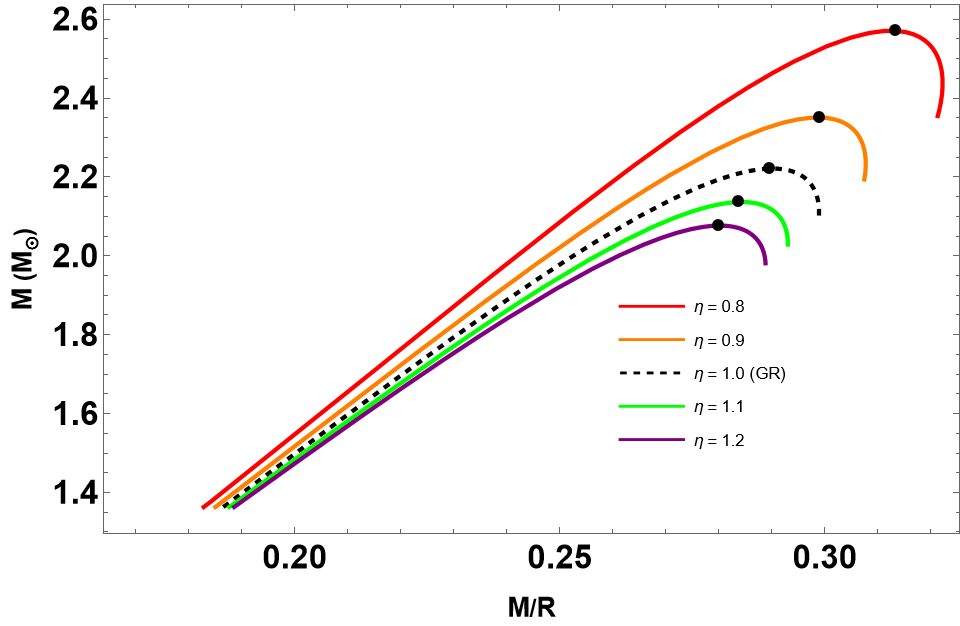}
    \caption{We display the mass-radius and mass-compactness relations for QSs in Rastall gravity. Here, we vary $\eta \in [0.8,1.2]$ and other chosen parameters are $B_{\rm eff} = 90$ MeV/fm$^3$ and $\bar{\lambda} = 0.6$, respectively. The horizontal bars represent the maximum mass constraints from PSR J0348+0432 (Light-Red) \cite{Antoniadis:2013pzd}, PSR J0740+6620 (Pink) \cite{Fonseca:2021wxt} and PSR J0952-0607 (Magenta) \cite{Romani:2022jhd}. Further imposing constraints from the gravitational wave signal GW190814 event (Purple) \cite{LIGOScientific:2020zkf}. The dashed-black line represents the curve for the GR solution. See text for more details.}
    \label{profiles_vary_eta1}
\end{figure}

\begin{figure}[h]
    \centering
    \includegraphics[width=8cm, height=6.3cm]{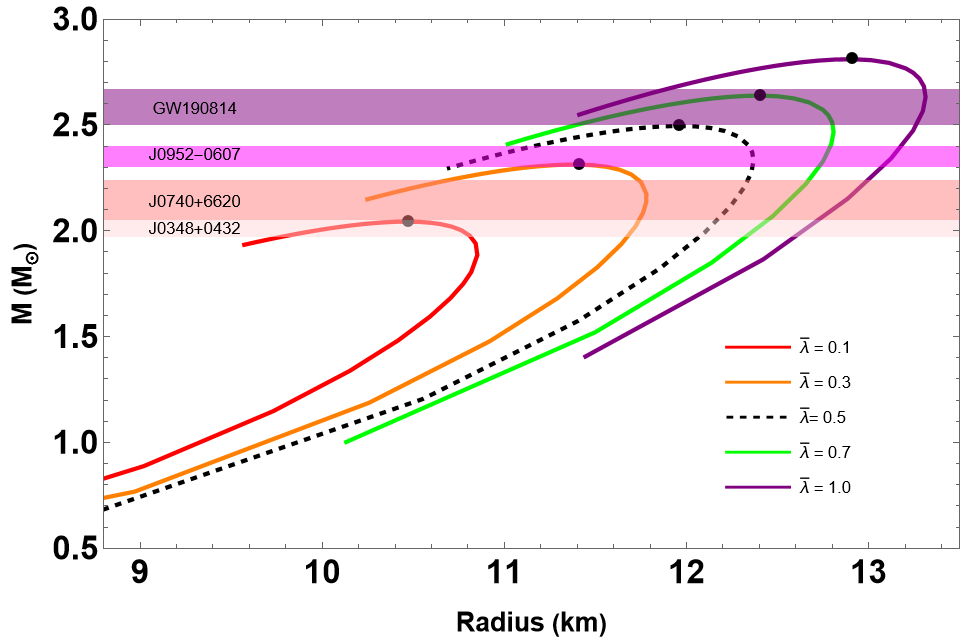}
    \includegraphics[width=8cm, height=6.3cm]{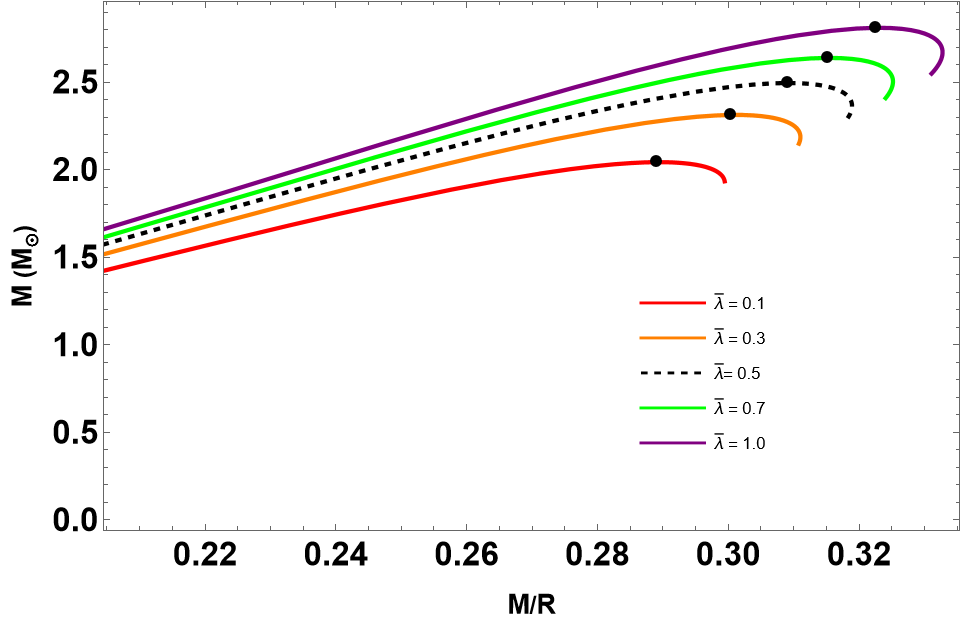}
    \caption{We display the mass-radius and mass-compactness relations for QSs in Rastall gravity. Here, we vary $\bar{\lambda} \in [0.1, 1.0]$, and other chosen parameters are $B_{\rm eff} = 90$ MeV/fm$^3$ and $\eta = 0.6$, respectively. To compare with observational constraints, we indicate the same pulsar measurements as in Fig. \ref{profiles_vary_eta1}.}
    \label{profiles_vary_eta2}
\end{figure}

\begin{figure}[h]
    \centering
    \includegraphics[width=8cm, height=6.3cm]{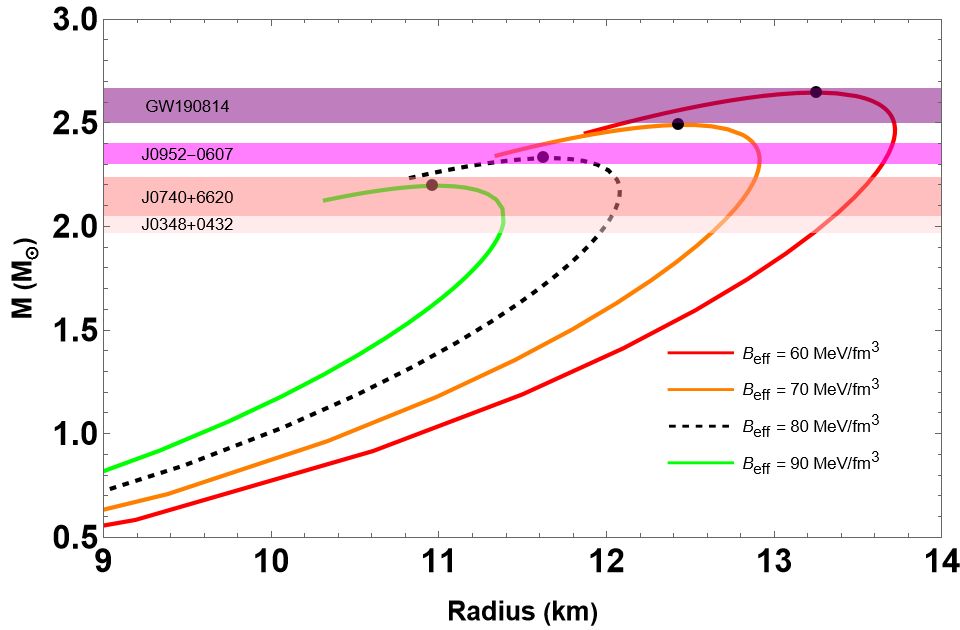}
    \includegraphics[width=8cm, height=6.3cm]{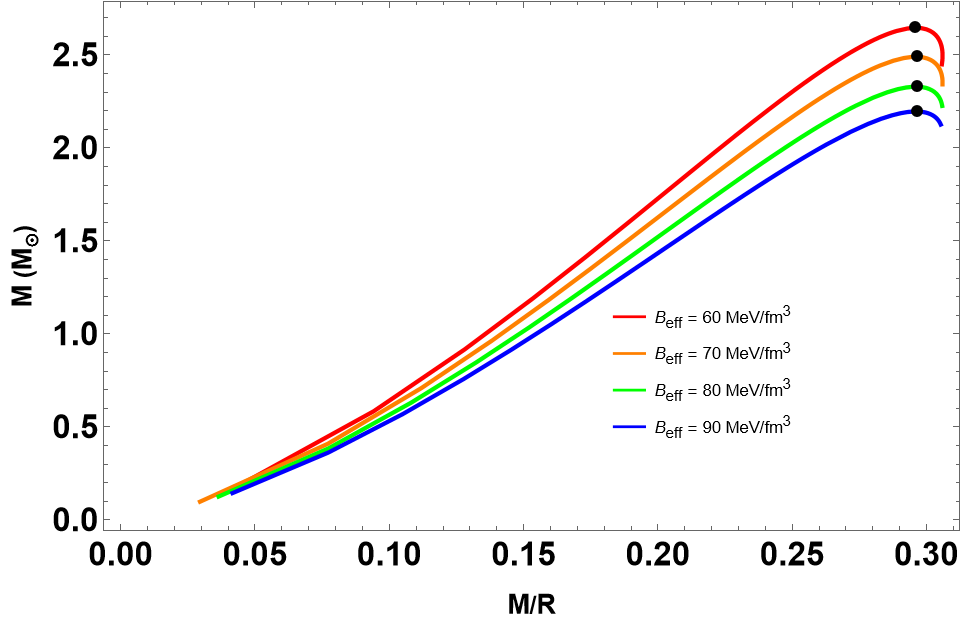}
    \caption{We display the mass-radius and mass-compactness relations for QSs in Rastall gravity. Here, we vary $B_{\rm eff} \in [60, 90]$ MeV/fm$^3$, and other chosen parameters are $\bar{\lambda} = 0.2$ and $\eta = 0.8$, respectively. To compare with observational constraints, we indicate the same pulsar measurements as in Fig. \ref{profiles_vary_eta1}.}
    \label{profiles_vary_B}
\end{figure}

\section{Numerical Results and Discussions}\label{sec4}

For the given EoS \eqref{eos_p}, we solve the hydrostatic equilibrium equations numerically and explore the effects of model parameters on the mass-radius relations of QSs. \textcolor{black}{We recover the conventional $(M, R)$ relation by introducing $(M, R) = (\bar{M}/\sqrt{4B_{\rm eff}}, \bar{R}/\sqrt{4B_{\rm eff}})$.} Additionally, we determine the stability of the configurations via the static stability criterion, the adiabatic index, and by checking the sound velocity.

We impose constraints from the GW190814 event by using the 2.6 $M_{\odot}$ mass of the secondary object as an upper bound to test the viability of QS models in Rastall gravity. A QS model is considered valid if it predicts a maximum mass at or above this threshold, which challenges traditional neutron star models based on general relativity. This constraint, combined with data from other massive NSs (e.g., PSR J0348+0432, PSR J0740+6620), helps validate the Rastall gravity framework for explaining such massive compact objects. In addition, in our analysis, we applied a QCD-inspired EoS throughout the entire structure of the quark star, including the crust. If a more traditional neutron star crust were considered, it would likely increase the radius and slightly affect the mass because of the additional matter. The transition between the crust and quark core could also modify the stability and mass-radius relation, though the core conclusions about the viability of quark stars in Rastall gravity, particularly their ability to achieve high masses, would remain robust. We acknowledge this possibility and plan to explore it in future work.

\subsection{Constraints on Model Parameters}

1. \textbf{Parameter $\eta$ (Rastall's Coupling Constant)}: We vary $\eta$ within the range [0.8, 1.2] as illustrated in Section \ref{case1} (see Fig. \ref{profiles_vary_eta1} and Table \ref{tableVaryEta}). By comparing the mass-radius relations derived from the model with observed data from massive pulsars, including PSR J0348+0432, PSR J0740+6620, and the compact object in the GW190814 event, we find that smaller values of $\eta$ (e.g., 0.8) yield maximum masses consistent with the observed upper bounds. Higher values of $\eta$ result in lower maximum masses, which are inconsistent with the astrophysical observations of heavy NSs. This is further supported by Maulana and Sulaksono's work \cite{isMaulana:2019sgd}, where they showed that the energy conditions restrict the value of $\eta$ for $\eta > 0$. In agreement with their findings, our results indicate that as $\eta$ decreases, the mass of quark stars increases, and the Rastall case allows for more massive quark stars compared to general relativity.

2. \textbf{Parameter $\bar{\lambda}$ (Interaction Parameter)}: As shown in Section \ref{case2} (Fig.  \ref{profiles_vary_eta2} and Table \ref{table2}), $\bar{\lambda}$ is varied between [0.1, 1.0]. The constraints on this parameter come from ensuring that the predicted maximum masses of quark stars match the observed masses of massive pulsars. Higher values of $\bar{\lambda}$, such as 1.0, correspond to more massive and larger-radius quark stars, fitting well within the range of observed pulsar data.

3. \textbf{Parameter $B_{\rm eff}$ (Effective Bag Constant)}: In Section \ref{case3} (Fig. \ref{profiles_vary_B} and Table \ref{tableVaryB}), we explore the effect of varying $B_{\rm eff}$ within the range [60, 90] MeV/fm$^3$. Our results indicate that increasing $B_{\rm eff}$ decreases both the maximum mass and the radius of the quark stars. This enables us to rule out values of $B_{\rm eff}$ that produce quark star masses below the 2 $M_{\odot}$ threshold observed for some pulsars, such as PSR J0348+0432.

In conclusion, the constraints on $\eta$, $\bar{\lambda}$, and $B_{\rm eff}$ were derived by varying these parameters and comparing the corresponding mass-radius relations with observational data. For the sake of clarity, we have outlined these constraints here. Moreover, in Sec. \ref{case1}, we have used $\eta = 0.6$ to explore the effects of varying the interaction parameter $\bar{\lambda}$ on QS properties within Rastall gravity. This value has been chosen to examine how $\bar{\lambda}$ influences the mass-radius relationship and stability when $\eta$ deviates from the standard $\eta = 1$ (corresponding to general relativity). While $\eta = 0.6$ has not been considered in Section IV A, we have included a discussion to compare the results with those for $\eta$ values within the range $[0.8, 1.2]$, as explored in the previous section.

\subsection{Profiles for Variation of $\eta$}\label{case1}

In this section, we present the structural properties of QSs, including mass-radius $(M-R)$ and mass-compactness $(M-M/R)$ relations, as shown in Fig. \ref{profiles_vary_eta1}. Here, we vary the Rastall parameter $\eta \in [0.8, 1.2]$, with other parameters set at $B_{\rm eff} = 90$ MeV/fm$^3$ and $\bar{\lambda} = 0.6$. To constrain the mass and radius relations, we compare our results with recent observational astrophysical data: PSR J0348+0432 with a mass of $M = 2.01 \pm 0.04 M_{\odot}$ (Light-Red) \cite{Antoniadis:2013pzd}, PSR J0740+6620 with a mass of $M = 2.08 \pm 0.07 M_{\odot}$ (Pink) \cite{Fonseca:2021wxt}, and PSR J0952-0607 with a mass of $M = 2.35 \pm 0.17 M_{\odot}$ (Magenta) \cite{Romani:2022jhd}. Additionally, we impose constraints from the gravitational wave event GW190814 (Purple) \cite{LIGOScientific:2020zkf}. As shown in Fig. \ref{profiles_vary_eta1}, the maximum mass increases with decreasing values of $\eta$. From our calculations, the maximum mass reaches up to 2.57 $M_{\odot}$ with a radius of 12.01 km for $\eta=0.8$, compatible with the mass of the secondary component in the GW190814 event. We also record maximum QS masses greater than 2 $M_{\odot}$, with the maximum mass for $\eta=1$ (GR case) being 2.22 $M_{\odot}$ with a radius of 11.37 km. This variation arises from the influence of $\eta$ on the $M-R$ relations (the dashed-black line in Fig. \ref{profiles_vary_eta1}). The $(M-M/R)$ curves in the lower panel of Fig. \ref{profiles_vary_eta1} show that the curves are almost indistinguishable at the low mass region. Additionally, the maximum compactness for $\eta = 1$ (GR solution) is $M/R = 0.313$, as indicated by the black-dashed curve. Details regarding the structural properties of QSs are presented in Table \ref{tableVaryEta}.

\begin{table}[h]
    \caption{We have tabulated the maximum mass $M_{\rm{max}}$ and the corresponding radius $R_{\rm{max}}$ in units of $M_\odot$ and km, respectively. The selected parameters are $B_{\rm eff} = 90$ MeV/fm$^3$ and $\bar{\lambda} = 0.6$, with variation in $\eta \in [0.8, 1.2]$.}
    \begin{ruledtabular}
    \begin{tabular}{ccccc}
    $\eta$  & $M$ [$M_\odot$]  & $R$ [km] & $\rho_c$ [MeV/fm$^3$] & $M/R$  \\
    \hline
        0.8 &  2.57  & 12.15 & 844 & 0.313 \\
        0.9 &  2.35 & 11.65 & 956 & 0.298 \\
        1.0 &  2.22  & 11.37 & 1013 & 0.289 \\
        1.1 &  2.13  & 11.16 & 1069 & 0.283 \\
        1.2 &  2.07  & 10.99 & 1125 & 0.280 
    \end{tabular}
    \end{ruledtabular}
    \label{tableVaryEta}
\end{table}

\subsection{Profiles for variation of \texorpdfstring{$\bar{\lambda}$}{l}} \label{case2}

 In Fig. \ref{profiles_vary_eta2}, we show the resulting $(M-R)$ and $(M-M/R)$ relationship in variation of $\bar{\lambda} \in [0.1, 1.0]$. We choose the other values for plotting are $B_{\rm eff} = 90$ MeV/fm$^3$ and $\eta = 0.6$, respectively. 
It should be pointed out that both the maximum mass and the corresponding radius of the QS increases with increasing values of $\bar{\lambda}$.  One can see from the Fig. \ref{profiles_vary_eta2} that the existence of QSs with masses  $> 2 M_{\odot}$ suggests the EoS is relatively stiff at high densities. Moreover, we can say that the considered EoS in this work support a larger maximum mass which is also compatible with the measured masses of pulsars same as of Fig. \ref{profiles_vary_eta1}.  In Table \ref{table2}, we show the main physical properties of QSs, where we have recorded the maximum gravitational mass is $2.81M_{\odot}$ with radius 12.91 km at $\bar{\lambda}=1.0$. Still, regarding the values shown in Table \ref{table2}, we remark that the maximum compactness lies
in the range of $0.289 < M/R < 0.322$. We have also noticed that the maximum compactness increases with 
increasing values of $\Bar{\lambda}$  and the results are presented in the lower panel of  Fig. \ref{profiles_vary_eta2}.

\begin{table}
\caption{\label{table2}  We have tabulated the maximum mass $M_{\rm{max}}$ and the corresponding radius $R_{\rm{max}}$ in units of $M_\odot$ and km, respectively. Selected parameters are $B_{\rm eff} = 90$ MeV/fm$^3$ and $\eta = 0.8$ $\bar{\lambda} = 0.6$ in variation of $\bar{\lambda} \in [0.1, 1.0]$.  }
\begin{ruledtabular}
\begin{tabular}{ccccc}
$\Bar{\lambda}$  & $M$ [$M_\odot$]  &  $R_{M}$  [\rm{km}]  & $\rho_c$ [MeV/fm$^3$] & $M/R$  \\
\colrule
0.1 &  2.04  &  10.47  &  1,069  &  0.289  \\
0.3 &  2.31  &  11.41  &  900  &  0.300  \\
0.5 &  2.50  &  11.96  &  844  &  0.309  \\
0.7 &  2.63  &  12.40  &  788  &  0.315  \\
1.0 &  2.81  &  12.91  &  731  &  0.322 
\end{tabular}
\end{ruledtabular}
\end{table}

\subsection{Profiles for variation of \texorpdfstring{$B_{\rm eff}$}{B}}\label{case3}

Finally, we end our discussion by varying the effective bag constant $B_{\rm eff}\in [60, 90]$ MeV/fm$^3$ in Fig. \ref{profiles_vary_B}. We set the other parameters are $\bar{\lambda} = 0.2$ and $\eta = 0.8$, respectively. From the Fig. \ref{profiles_vary_B}, it is clear 
that the maximum mass decreases with increasing values of $B_{\rm eff}$. Some quantities related to the maximum mass and the corresponding radius are shown in Table \ref{tableVaryB}. Depending on the strength of the bag constant, it is found that the massive QSs exits with masses 2.20 to 2.65$M_{\odot}$. The radii of all these stars are between 10-12 km, as expected for compact  stars.  We next explore the maximum compactness vs mass of the QS i.e., $(M-M/R)$ relation in the lower panel of Fig. \ref{profiles_vary_B}. It can be seen that the maximum compactness does not affected by increasing or decreasing values of $B_{\rm eff}$.  Furthermore,  we gather the data for maximum compactness in Table \ref{tableVaryB} depending on the parameters and is found to be $0.297$.

\begin{figure}[h]
    \centering
     \includegraphics[width = 8 cm, height=6.3cm]{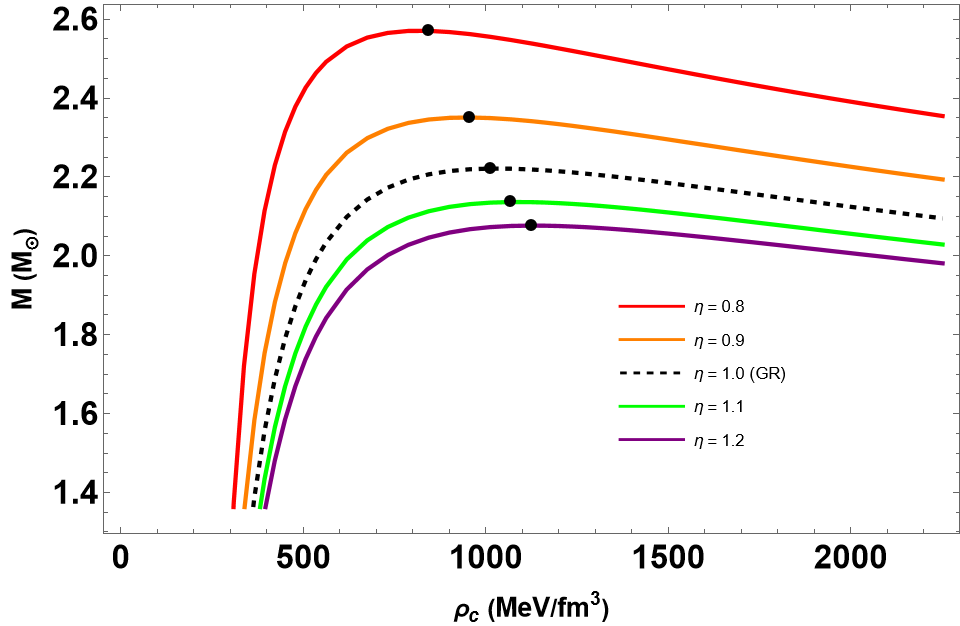}
     \includegraphics[width = 8 cm, height=6.3cm]{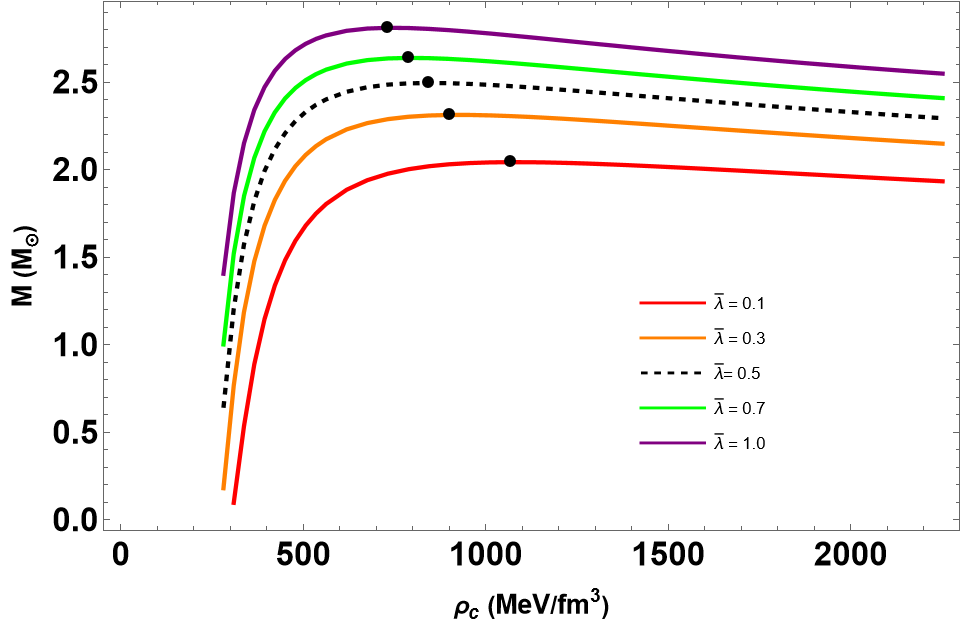}
     \includegraphics[width = 8 cm, height=6.3cm]{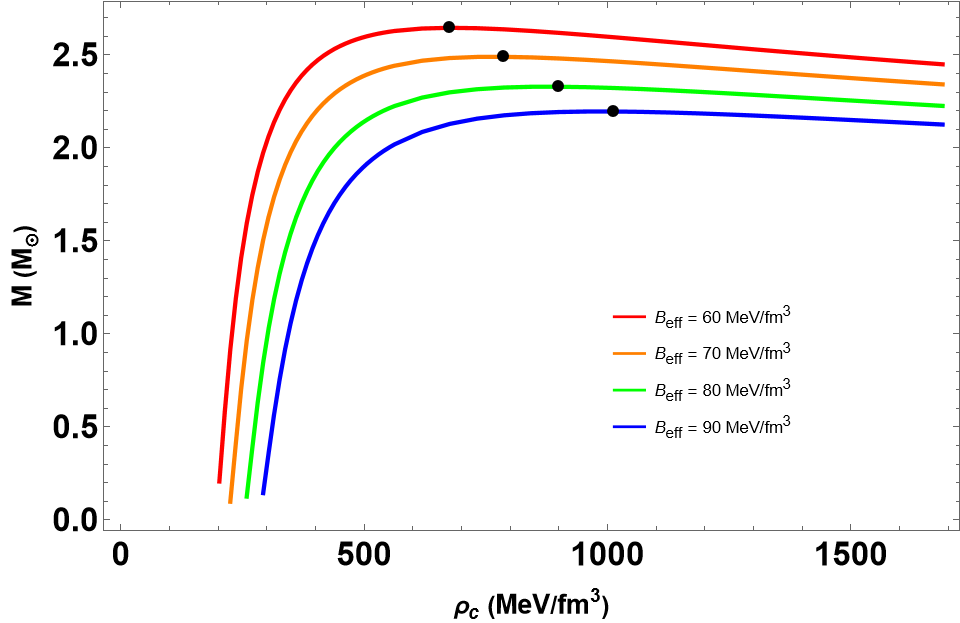}
    \caption{Profiles of the mass versus central density relations. The parameters used are the same as in Figs. \ref{profiles_vary_eta1} to \ref{profiles_vary_B}.}
    \label{profiles_vary_stc}
\end{figure}

\begin{figure}[H]
    \centering
     \includegraphics[width = 8 cm, height=6.3cm]{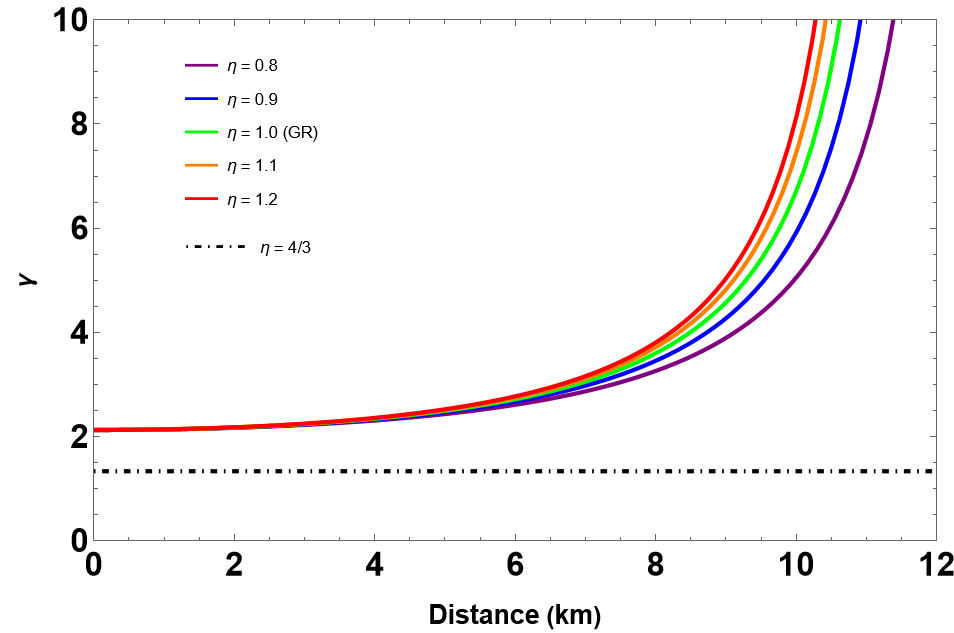}
     \includegraphics[width = 8 cm, height=6.3cm]{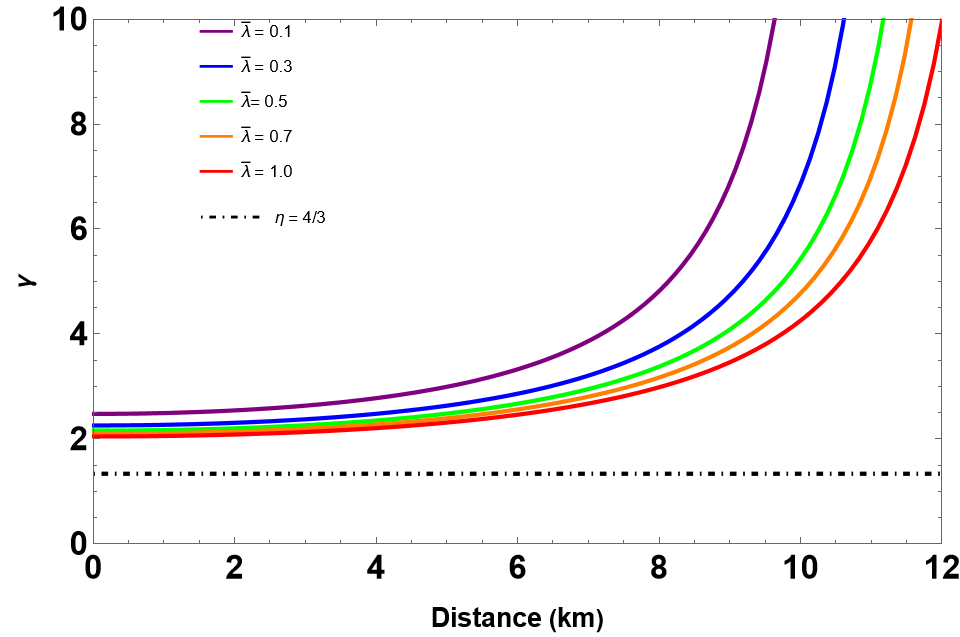}
     \includegraphics[width = 8 cm, height=6.3cm]{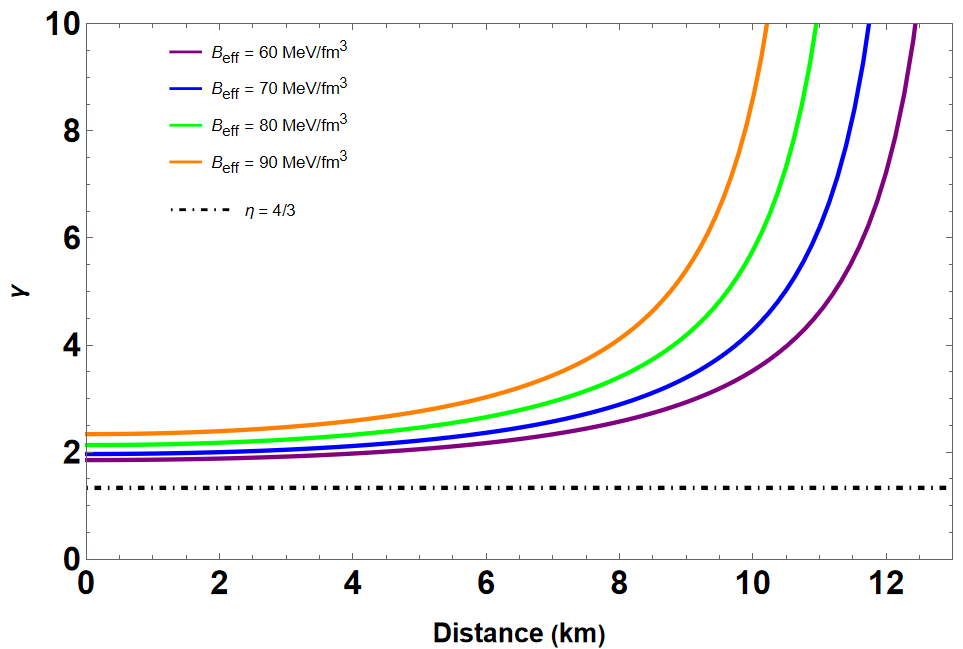}
    \caption{Graphs of adiabatic index for QSs, computed using Eq. (\ref{eos_p}). The parameters considered are the same as those in Figs. \ref{profiles_vary_eta1} to \ref{profiles_vary_B}.}
    \label{profiles_vary_gamma}
\end{figure}

\begin{figure}[H]
    \centering
     \includegraphics[width = 8 cm, height=6.3cm]{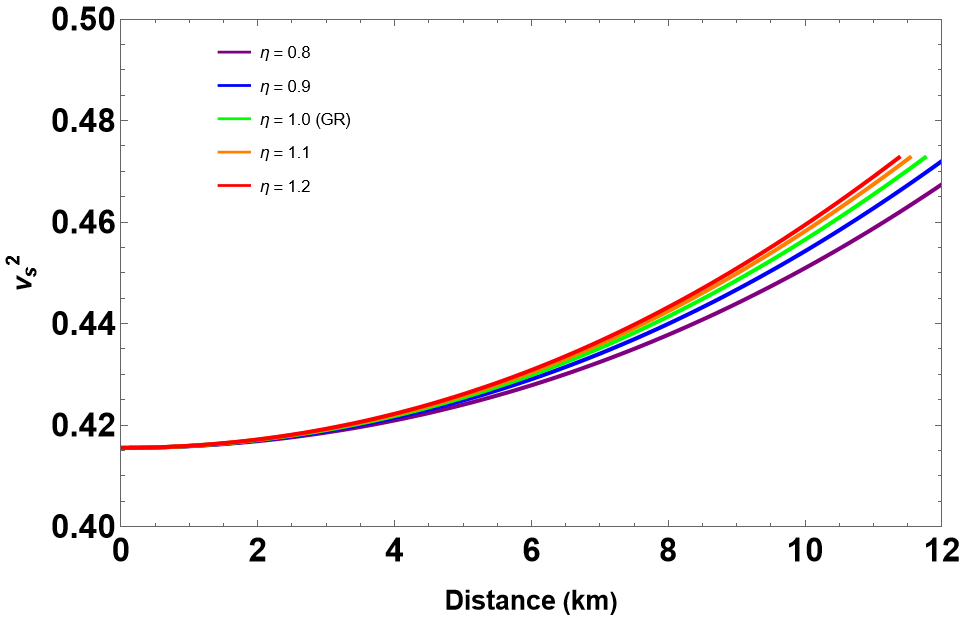}
     \includegraphics[width = 8 cm, height=6.3cm]{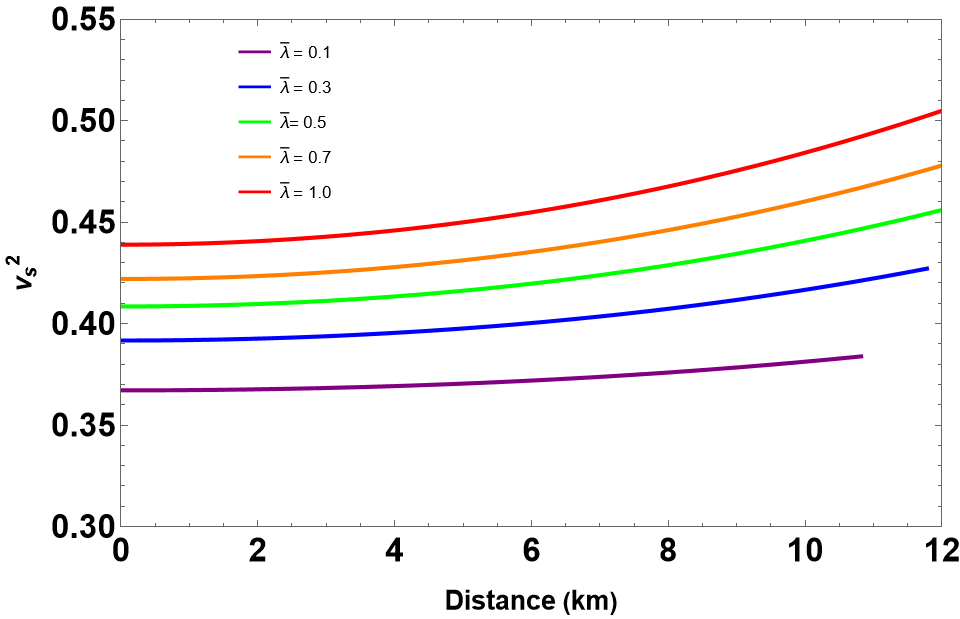}
     \includegraphics[width = 8 cm, height=6.3cm]{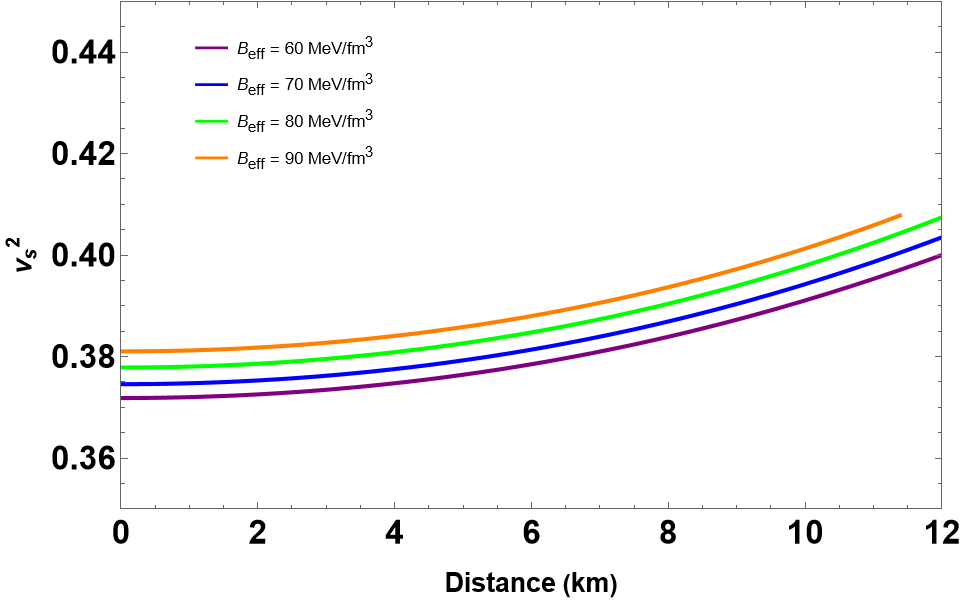}
    \caption{The squared speed of sound $v^2_s = dp/d\rho$ plotted as a function of the radial coordinate, $r$. \textcolor{black}{The parameters used are the same as those in} Figs. \ref{profiles_vary_eta1} to \ref{profiles_vary_B}.}
    \label{profiles_vary_velocity}
\end{figure}

\begin{table}[H]
    \caption{We have tabulated the maximum mass $M_{\rm{max}}$ and the corresponding radius $R_{\rm{max}}$ in units of $M_\odot$ and km, respectively. Selected parameters are $\bar{\lambda} = 0.2$ and $\eta = 0.8$ in variation of $B_{\rm eff}\in [60, 90]$ MeV/fm$^3$.}
    \begin{ruledtabular}
    \begin{tabular}{ccccc}
    $B_{\rm eff}$  & $M$   &  $R_{M}$  & $\rho_c$  & $M/R$  \\
    MeV/fm$^3$ & $M_{\odot}$ &   km & MeV/fm$^3$ & \\
    \hline
        60 &  2.65  & 13.25 & 1013 & 0.297 \\
        70 &  2.49  & 12.43 & 900 & 0.297 \\
        80 &  2.32  & 11.63 & 788 & 0.297 \\
        90 &  2.20  & 10.96 & 675 & 0.296
    \end{tabular}
    \end{ruledtabular}
    \label{tableVaryB}
\end{table}

\section{Equilibrium and Stability of \texorpdfstring{QS\lowercase{s}}{QS}}\label{sec5}

To evaluate the stability of QSs in Rastall gravity, we employed the static stability criterion, adiabatic index, and sound velocity. These measures help us ascertain the stability of the stars under study.

Although Chandrasekhar's stability criterion \cite{isDu:2006kg} was developed within the framework of general relativity, its application in Rastall gravity remains valid. The criterion is fundamentally based on the star's response to radial perturbations, which is governed by hydrostatic equilibrium, not specific to GR alone. Recent studies (e.g., Refs. \cite{isPretel:2024lae,isMaulana:2019sgd}) have shown that stability conditions involving the adiabatic index $\gamma > 4/3$ are still applicable in modified gravity theories, including Rastall gravity. Thus, we confidently apply this criterion in our analysis, recognizing that the specific dynamics of Rastall gravity might introduce slight modifications. However, while radial oscillations are crucial for a complete stability analysis, we chose to focus on simpler stability criteria such as static stability, adiabatic index, and causality conditions in this study. Analyzing radial oscillations, which requires solving the Sturm-Liouville eigenvalue problem, would expand the scope of this paper beyond our primary goal of exploring the basic structural properties of QSs in Rastall gravity. However, we acknowledge the importance of this criterion and plan to include a detailed analysis of radial oscillations in future work to complement our current findings.

\subsection{Static Stability Criterion}

The study of radial oscillation modes offers a promising method for assessing the stability of compact stars. We focus on the \textit{static stability criterion} \cite{harrison,ZN}, plotting the total mass versus central density, i.e., $M$ vs. $\rho_c$, for QSs using Eq. (\ref{eos_p}). Previous studies, such as those by \cite{isMaulana:2019sgd}, have examined the stability of QSs in Rastall gravity. It is essential to note that, while this is a necessary condition for stability, it is not sufficient. The condition can be expressed by the following inequalities:
\begin{eqnarray}
\frac{dM}{d\rho_c} < 0 &~ \rightarrow \text{indicates an unstable configuration}, \\
\frac{dM}{d\rho_c} > 0 &~ \rightarrow \text{indicates a stable configuration}.
\label{criterion_M_rho_c}
\end{eqnarray}
In Fig. \ref{profiles_vary_gamma}, we illustrate several $M-\rho_c$ graphs with parameters as given in Figs. \ref{profiles_vary_eta1} to \ref{profiles_vary_B}. In these figures, the turning points (marked by a black circle) delineate the boundary separating the stable configuration region ($dM/d\rho_c > 0$) from the unstable one.

\subsection{Adiabatic Indices}

Another critical aspect to examine is the adiabatic index $\gamma$, which is crucial to assess the stability of the configuration. We calculate $\gamma$ using the method developed by Chandrasekhar \cite{Chandrasekhar}, defined by the following equation:
\begin{eqnarray}\label{adi}
    \gamma \equiv \left(1+\frac{\rho}{p}\right)\left(\frac{dp}{d\rho}\right)_S,
\end{eqnarray}
where $dp/d\rho$ is the speed of sound and the subscript $S$ indicates derivation at constant entropy. For isotropic fluid spheres, $\gamma$ must satisfy certain conditions that govern the dynamical instability of relativistic objects. According to \cite{Glass}, the stability domain is defined by the condition $\gamma > \gamma_{cr} = 4/3$ for polytropic spheres, where $\gamma_{cr}$ is the critical adiabatic index. Thus, the stability condition can be established as $\gamma > 4/3$, which was further examined by Moustakidis \cite{Moustakidis:2016ndw} for the Tolman VII solution. We present our results for the adiabatic index $\gamma$ in Fig. \ref{profiles_vary_gamma}, which indicate that our considered model is dynamically stable.

\subsection{Sound Speed and Causality}

We now investigate the stability of configurations by examining the sound velocity, determined by the relation $v^2_s = dp/d\rho$. According to the causality condition, $v^2_s < 1$, meaning the speed of sound should be less than the speed of light. Using the given EoS \eqref{eos_p} and our numerical results, we display the propagation of sound within the stellar radius in Fig. \ref{profiles_vary_velocity}. As shown, the proposed model for QS is stable within the framework of Rastall theory.

While the causality condition (i.e., $v_s^2 < 1$) ensures that the equation of state is physically reasonable, it is not the sole basis for our stability conclusions. The overall stability was derived from a combination of criteria, including the adiabatic index $\gamma > 4/3$, which ensures dynamical stability, and the static stability criterion ($dM/d\rho_c > 0$). These combined analyses allow us to confidently conclude that the proposed QS models are stable within the framework of Rastall gravity. In Figs. \ref{profiles_vary_gamma} and \ref{profiles_vary_velocity}, we present the radial profiles of the adiabatic index and the speed of sound for the QS models discussed in Sections IV.A, IV.B, and IV.C. Specifically, Fig. \ref{profiles_vary_gamma} shows the adiabatic index profiles for varying values of $\eta$, $\bar{\lambda}$, and $B_{\rm eff}$, while Fig. \ref{profiles_vary_velocity} illustrates the corresponding speed of the sound profiles for the same parameter variations. The models cover ranges of $\eta \in [0.8, 1.2]$, $\bar{\lambda} \in [0.1, 1.0]$, and $B_{\rm eff} \in [60, 90]$ MeV/fm$^3$, with the other parameters fixed as appropriate in each case. These profiles demonstrate how varying these key parameters influences the internal structure and stability of QSs within Rastall gravity. 

{\color{black}Additionally, it should be noted that the squared speed of sound $v_s^2 = \frac{dp}{d\rho}$ in Fig. \ref{profiles_vary_velocity} shows a slight divergence between the "brinjal" (purple) and red lines in the 11 to 12 km range, reflecting the sensitivity of the model to parameter variations such as the Rastall parameter $\eta$, interaction parameter $\bar{\lambda}$, and effective bag constant $B_{\rm eff}$. This divergence suggests that the equation of state (EoS) becomes stiffer in the outer layers, where density gradients are smaller. The dotted lines in the figure represent the GR case ($\eta = 1$), included as a baseline for comparison, illustrating the distinct effects of Rastall gravity on the sound speed profile, particularly in the outer regions of the star.}

\section{Final Remarks and Future Directions}\label{sec6}

In this article, we perform a comprehensive study of compact stars composed of strange quark matter. The assumed EoS is based on the newly proposed IQM model, which includes effects from pQCD corrections and color superconductivity \cite{isZhang:2020jmb}. Specifically, using this EoS, we numerically solved the modified TOV equations for spherically symmetric configurations within the framework of Rastall gravity. This theory, which generalizes the conservation principles of the stress-energy tensor to allow for $T^{\mu}_{\nu;\mu} \neq 0$ in curved spacetime, argues in favor of $T^{\mu}_{\nu;\mu} \propto R_{;\nu}$, i.e., the covariant divergence of the stress-energy tensor is proportional to the gradient of the Ricci scalar. Rastall theory is a straightforward generalization of GR that has been successfully tested in strong gravity fields.

We focused on the physical properties of stars, such as mass-radius and compactness relations, and analyzed their stability depending on model parameters ($\bar{\lambda}, \eta, B_{\rm eff}$). We primarily examined the influence of the Rastall parameter $\eta$ on the $M-R$ relations, finding that for $\eta > 0$, the maximum mass increases with decreasing $\eta$ values, as shown in Fig. \ref{profiles_vary_eta1}. Conversely, in cases where $\bar{\lambda} > 0$, we observed that the maximum mass increases with increasing $\bar{\lambda}$ values, reaching up to 2.81$M_{\odot}$ with a radius of 12.91 km for $\bar{\lambda}=1.0$ (see Fig. \ref{profiles_vary_eta2}). Additionally, we described the effects of varying the bag constant $B_{\rm eff}$, noting that the maximum mass decreases with increasing $B_{\rm eff}$ values (see Fig. \ref{profiles_vary_B}). Our results indicate that the maximum mass of QSs in Rastall gravity can exceed the 2$M_{\odot}$ limit, consistent with observational constraints from measurements of heavy NSs. However, exploring a broader region of the parameter space, including extreme values of $\eta$, $\bar{\lambda}$, and $B_{\rm eff}$, could reveal critical thresholds and more pronounced deviations in the QS properties. For example, extreme $\eta$ values may lead to highly compact or excessively large stars, while very high $\lambda$ could stiffen the EoS, resulting in higher maximum masses. In future work, we plan to expand our analysis to cover more of the parameter space, including simultaneous variations, to detect potential patterns and critical behavior.

In conclusion, our study demonstrates that QSs in Rastall gravity can achieve higher maximum masses compared to GR, making them consistent with observations of massive pulsars, such as those seen in the GW190814 event. By examining stability criteria and comparing theoretical predictions with observational data, we constrained the parameters $\eta$, $\bar{\lambda}$, and $B_{\rm eff}$, confirming that Rastall gravity provides a viable framework for describing stable, massive QSs. These findings suggest that Rastall gravity could serve as a promising alternative to general relativity for modeling compact stars in extreme gravitational environments.

Furthermore, we addressed the stability of these stars by applying the static stability criterion, examining the adiabatic index, and checking the sound velocity. \textcolor{black}{Overall, our study demonstrates that stable QSs can exist with higher masses than those predicted by their counterparts in GR.} Moving forward, we aim to extend our analysis to include more diverse conditions and explore additional modified gravitational physics that may provide further insights into the behavior and characteristics of compact stars under extreme conditions.
\\

\begin{acknowledgments} \textcolor{black}{We would like to express our sincere gratitude to the Editor and Reviewers for their insightful comments and valuable suggestions, which have greatly contributed to improving the quality and clarity of our manuscript.} A.~Pradhan expresses gratitude to the IUCCA in Pune, India, for offering facilities under associateship programs. \.{I}.~S. expresses gratitude to T\"{U}B\.{I}TAK, ANKOS, and SCOAP3 for their financial support. He also acknowledges COST Actions CA22113 and CA21106 for their contributions to networking.

\end{acknowledgments}\


\begin{thebibliography}{90}

\bibitem{isZhang:2024xod}
X.~L.~Zhang, Y.~F.~Huang and Z.~C.~Zou,
[arXiv:2404.00363 [astro-ph.HE]].


\bibitem{isSu:2024znh}
L.~Q.~Su, C.~Shi, Y.~F.~Huang, Y.~Yan, C.~M.~Li, W.~L.~Yuan and H.~S.~Zong,
Astrophys. Space Sci. \textbf{369},  29 (2024).


\bibitem{isKolomeitsev:2024gek}
E.~E.~Kolomeitsev and D.~N.~Voskresensky,
[arXiv:2404.09875 [nucl-th]].

\bibitem{isPretel:2024lae}
J.~M.~Z.~Pretel and C.~E.~Mota,
Gen. Rel. Grav. \textbf{56},  43 (2024).


\bibitem{isSen:2024reu}
D.~Sen, H.~Gil and C.~H.~Hyun,
[arXiv:2402.18912 [nucl-th]].


\bibitem{isFarrell:2024bka}
D.~Farrell, F.~Weber, M.~G.~Orsaria, \textit{et al.}
[arXiv:2402.08835 [astro-ph.HE]].

\bibitem{isTANGPHATI2024}
T.~Tangphati, A.~Banerjee, \.I.~Sakall\i{}, and A.~Pradhan,
Chin. J. Phys. \textbf{90}, 422 (2024). 

\bibitem{isBaym:2017whm}
G.~Baym, T.~Hatsuda, T.~Kojo, P.~D.~Powell, Y.~Song and T.~Takatsuka,
Rept. Prog. Phys. \textbf{81},  056902 (2018).

\bibitem{isLOFT:2011pkp}
M.~Feroci \textit{et al.} [LOFT],
Exper. Astron. \textbf{34}, 415 (2012).

\bibitem{isDarabi:2017coc}
F.~Darabi, H.~Moradpour, I.~Licata, Y.~Heydarzade and C.~Corda,
Eur. Phys. J. C \textbf{78}, 25 (2018).


\bibitem{isOliveira:2015lka}
A.~M.~Oliveira, H.~E.~S.~Velten, J.~C.~Fabris and L.~Casarini,
Phys. Rev. D \textbf{92},  044020 (2015).

\bibitem{isHeydarzade:2016zof}
Y.~Heydarzade, H.~Moradpour and F.~Darabi,
Can. J. Phys. \textbf{95}, 1253 (2017).

\bibitem{isFabris:2012hw}
J.~C.~Fabris, O.~F.~Piattella, D.~C.~Rodrigues, C.~E.~M.~Batista and M.~H.~Daouda,
Int. J. Mod. Phys. Conf. Ser. \textbf{18}, 67 (2012).

\bibitem{isOvgun:2019jdo}
A.~\"Ovg\"un, \.I.~Sakall\i{}, J.~Saavedra and C.~Leiva,
Mod. Phys. Lett. A \textbf{35}, 2050163 (2020).

\bibitem{isDeMoraes:2019mef}
W.~A.~G.~De Moraes and A.~F.~Santos,
Gen. Rel. Grav. \textbf{51},  167 (2019).

\bibitem{isAnnala:2017tqz}
E.~Annala, C.~Ecker, C.~Hoyos, N.~Jokela, D.~Rodr\'\i{}guez Fern\'andez and A.~Vuorinen,
JHEP \textbf{12}, 078 (2018).

\bibitem{isKurkela:2014vha}
A.~Kurkela, E.~S.~Fraga, J.~Schaffner-Bielich and A.~Vuorinen,
Astrophys. J. \textbf{789}, 127 (2014).

\bibitem{isMariani:2019vve}
M.~Mariani, M.~G.~Orsaria, I.~F.~Ranea-Sandoval and G.~Lugones,
Mon. Not. Roy. Astron. Soc. \textbf{489},  4261 (2019).

\bibitem{isFreedman:1977gz}
B.~Freedman and L.~D.~McLerran,
Phys. Rev. D \textbf{17}, 1109 (1978).

\bibitem{isHasenfratz:1977dt}
P.~Hasenfratz and J.~Kuti,
Phys. Rept. \textbf{40}, 75 (1978).


\bibitem{isBaym:1976yu}
G.~Baym and S.~A.~Chin,
Phys. Lett. B \textbf{62}, 241 (1976).

\bibitem{isReya:1979zk}
E.~Reya,
Phys. Rept. \textbf{69}, 195 (1981).

\bibitem{isGross:2022hyw}
F.~Gross, E.~Klempt, S.~J.~Brodsky,  \textit{et al.}
Eur. Phys. J. C \textbf{83}, 1125 (2023).

\bibitem{isBuballa:2003qv}
M.~Buballa,
Phys. Rept. \textbf{407}, 205 (2005).

\bibitem{isCollins:1974ky}
J.~C.~Collins and M.~J.~Perry,
Phys. Rev. Lett. \textbf{34}, 1353 (1975).

\bibitem{isAnnala:2019puf}
E.~Annala, T.~Gorda, A.~Kurkela, J.~N\"attil\"a and A.~Vuorinen,
Nature Phys. \textbf{16},  907 (2020).

\bibitem{isWeissenborn:2011qu}
S.~Weissenborn, I.~Sagert, G.~Pagliara, \textit{et al.}
Astrophys. J. Lett. \textbf{740}, L14 (2011).

\bibitem{isMiller:2021qha}
M.~C.~Miller, F.~K.~Lamb, A.~J.~Dittmann,  \textit{et al.}
Astrophys. J. Lett. \textbf{918},  L28 (2021).

\bibitem{isMcLerran:2007qj}
L.~McLerran and R.~D.~Pisarski,
Nucl. Phys. A \textbf{796}, 83 (2007).


\bibitem{isGhaemmaghami:2023sho}
S.~A.~Ghaemmaghami and M.~Ghazanfari Mojarrad,
Eur. Phys. J. Plus \textbf{138},  970 (2023).

\bibitem{isMishra:2024jmc}
A.~N.~Mishra, G.~Pai\'c, C.~V.~Pajares, R.~P.~Scharenberg and B.~K.~Srivastava,
Universe \textbf{10},  55 (2024).

\bibitem{isShankaranarayanan:2022wbx}
S.~Shankaranarayanan and J.~P.~Johnson,
Gen. Rel. Grav. \textbf{54},  44 (2022).

\bibitem{isYuan:2016pkz}
F.~F.~Yuan and P.~Huang,
Class. Quant. Grav. \textbf{34},  077001 (2017).

\bibitem{isAntoniadis:2013pzd}
J.~Antoniadis, P.~C.~C.~Freire, N.~Wex,  \textit{et al.}
Science \textbf{340}, 6131 (2013).

\bibitem{isCardoso:2019rvt}
V.~Cardoso and P.~Pani,
Living Rev. Rel. \textbf{22}, 4 (2019).

\bibitem{isBaiotti:2016qnr}
L.~Baiotti and L.~Rezzolla,
Rept. Prog. Phys. \textbf{80},  096901 (2017).

\bibitem{issakalli:2022xrb}
\.I.~Sakalli and S.~Kanzi,
Turk. J. Phys. \textbf{46},  51 (2022).


\bibitem{PhysRevD109063008}
Y.~Yang, C.~Wu, and J.-F.~Yang,
Phys. Rev. D \textbf{109}, 063008 (2024).

\bibitem{EurPhysJA5620}
F.~Wu and C.~Wu,
Eur. Phys. J. A \textbf{56}, 20 (2020).

\bibitem{PhysRevC105015807}
F.~Ma, W.~Guo and C.~Wu,
Phys. Rev. C \textbf{105}, 015807 (2022)

\bibitem{PhysRevC107045804}
F.~Ma, C.~Wu and W.~Guo,
Phys. Rev. C \textbf{107}, 045804 (2023)


\bibitem{isTangphati:2024war}
T.~Tangphati, I.~Sakalli, A.~Banerjee and A.~Pradhan,
[arXiv:2404.01970 [gr-qc]].

\bibitem{isMaulana2019}
H.~Maulana and A.~Sulaksono,
Phys. Rev. D \textbf{100}, 124014 (2019).

\bibitem{isZhang:2020jmb}
C.~Zhang and R.~B.~Mann,
Phys. Rev. D \textbf{103}, 063018 (2021). 

\bibitem{isZhang:2021fla}
C.~Zhang,
Phys. Rev. D \textbf{104},  083032 (2021).

\bibitem{isRastall:1972swe}
P.~Rastall,
Phys. Rev. D \textbf{6}, 3357 (1972).

\bibitem{isRastall:1976uh}
P.~Rastall,
Can. J. Phys. \textbf{54}, 66 (1976).


\bibitem{Velten:2016bdk}
H.~Velten, A.~M.~Oliveira and A.~Wojnar,
PoS \textbf{MPCS2015}, 025 (2016).

\bibitem{Maulana:2019}
H.~Maulana, and A.~Sulaksono, 
Phys. Rev. D \textbf{100}, 124014 (2019).


\bibitem{Antoniadis:2013pzd}
J.~Antoniadis, P.~C.~C.~Freire, N.~Wex,  \textit{et al.}
Science \textbf{340}, 6131 (2013). 

\bibitem{Fonseca:2021wxt}
E.~Fonseca, H.~T.~Cromartie, T.~T.~Pennucci, \textit{et al.}
Astrophys. J. Lett. \textbf{915}, L12 (2021).

\bibitem{Romani:2022jhd}
R.~W.~Romani, D.~Kandel, A.~V.~Filippenko, \textit{et al.}
Astrophys. J. Lett. \textbf{934}, L17 (2022).

\bibitem{LIGOScientific:2020zkf}
R.~Abbott \textit{et al.} [LIGO Scientific and Virgo],
Astrophys. J. Lett. \textbf{896}, L44 (2020).

\bibitem{isMaulana:2019sgd}
H.~Maulana and A.~Sulaksono,
Phys. Rev. D \textbf{100}, 124014 (2019). 

\bibitem{isDu:2006kg}
J.~l.~Du,
New Astron. \textbf{12}, 60 (2006).


\bibitem{harrison} 
B.~K.~Harrison, 
\textit{Gravitational Theory and Gravitational Collapse}, University of Chicago Press, Chicago, 1965.

\bibitem{ZN} 
Y.~B.~Zeldovich, and I.~D.~Novikov, 
\textit{Relativistic Astrophysics, Vol.~I: Stars and Relativity}, University of Chicago Press, Chicago, 1971.


\bibitem{Chandrasekhar}
S. Chandrasekhar,  Astrophys. J. {\bf 140}, 417 (1964).

\bibitem{Glass}
E. N. Glass and A. Harpaz,  Mon. Not. Roy. Astron. Soc., {\bf 202},  1 (1983).

\bibitem{Moustakidis:2016ndw}
C.~C.~Moustakidis,
Gen. Rel. Grav. \textbf{49}, 68 (2017).


\end{thebibliography}
\end{document}